\documentclass[pra,reprint,amsmath,amssymb,aps,]{revtex4-2}

\usepackage{subfiles}
\usepackage{graphicx}
\usepackage{dcolumn}
\usepackage{hyperref}
\usepackage[caption=false]{subfig}
\usepackage[export]{adjustbox}
\usepackage{xcolor}
\usepackage[utf8]{inputenc}
\usepackage[T1]{fontenc}
\usepackage{siunitx}
\usepackage[acronym]{glossaries}

\providecommand{\vect}[1]{{\boldsymbol{#1}}}

\begin{document}

\title{Leveraging Interactions for Efficient Swarm-Based Brownian Computing}

\author{Alessandro Pignedoli}
\affiliation{
 Faculty of Physics and Center for Nanointegration Duisburg-Essen (CENIDE), University of Duisburg-Essen, 47057 Duisburg, Germany
}

\author{Atreya Majumdar}
\affiliation{
 Faculty of Physics and Center for Nanointegration Duisburg-Essen (CENIDE), University of Duisburg-Essen, 47057 Duisburg, Germany
}

\author{Karin Everschor-Sitte}
\affiliation{
Faculty of Physics and Center for Nanointegration Duisburg-Essen (CENIDE), University of Duisburg-Essen, 47057 Duisburg, Germany
}

\date{\today}

\begin{abstract}
Drawing inspiration from swarm intelligence, we show that short-range attractive interactions between thermally driven Brownian quasiparticles enable energy-efficient optimization.
As quasiparticles can be generated directly within a material, the swarm size can be adjusted with minimal energy overhead.
Using an optimization task defined by a spatially varying temperature landscape, we quantitatively show that interacting swarms reliably identify global optima and significantly outperform non-interacting searchers within a well-defined regime of interaction strength and swarm size. This improvement arises from emergent cooperative behavior, where local interactions guide the swarm toward high-quality solutions without central coordination. To link our physical model to experimental realizations, we coarse-grain the quasiparticle dynamics onto a sensor lattice and generate trajectories emulating particle-tracking measurements. 
We further show that the interacting swarm adapts robustly to landscapes that evolve over time.
These findings establish interacting Brownian quasiparticles as a physical platform for scalable and energy-efficient unconventional computing.
\end{abstract}

\maketitle

\section{Introduction}
Research on swarm intelligence inspired by biological systems has shown that large groups of simple agents, each following only local interaction rules, can collectively perform complex tasks, such as efficient resource localization, robust decision-making, and adaptive responses to changing environments \cite{Bonabeau1999, Hinchey2007, Rubenstein2014, Patnaik2017, Kaspar2021, Nguyen2024}.

A representative example is the Beeclust algorithm, where local agent-agent interactions promote clustering to solve optimization tasks~\cite{Kernbach2009,
Schmickl2011, Bodi2012, Kernbach2013, Heinrich2022, Rios2025}

In a different research field, Brownian (quasi-)particles explore the state space of a physical system through thermally driven motion independent of external stimuli~\cite{Brown1828}.
Such dynamics have recently attracted interest as a platform for thermometry~\cite{Chung2009, Geiss2020, EverschorSitte2024DE} as well as energy-efficient information processing, often referred to as Brownian computing~\cite{Norton2013, Peper2013, Goto2021, Raab2022, EverschorSitte2023DE, Beneke2024}.

\begin{figure}
    \centering
\includegraphics[width=0.9\linewidth]{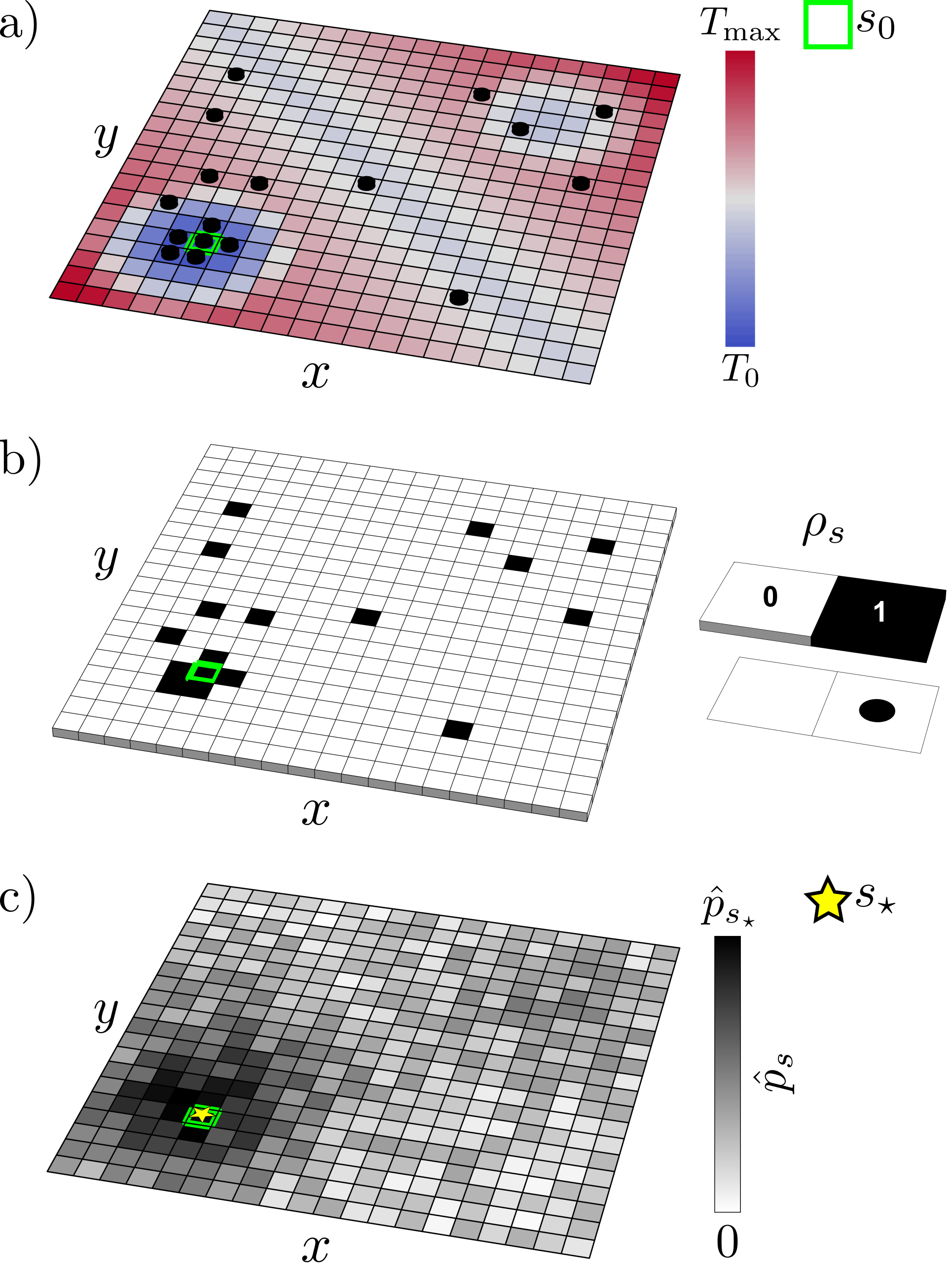}
    \caption{
    Schematic representation of swarm-based Brownian computing. a) Snapshot of interacting Brownian quasiparticles diffusing toward the global optimum $s_0$, indicated by the green square, within the temperature landscape shown in color.
    b) Discrete, coarse-grained representation of the system configuration $\vect{\rho}$ at a fixed time. 
    c) Single-particle probability distribution $\hat{\vect{p}}$, obtained from the time-averaged occupation vector $\vect{\rho}(t)$. The yellow star indicates the position $s_*$ of the statistical mode of the probability distribution.
    The system is confined to a square domain of size $L\times L$, and discretized into $M=\ell^2$ sensors. 
    }
    \label{fig:1}
\end{figure}

Here, we bring these two perspectives together by realizing the swarm agents as Brownian quasiparticles~\cite{EverschorSitte2023DE, EverschorSitte2024DE}.
 In contrast to engineered swarms, where scaling the number of agents incurs substantial fabrication and control overhead~\cite{Norton2013, Zhu2024, Nitti2025, teVrugt2026, Alhafnawi2026}, 
 quasiparticles can be generated within the material itself with minimal additional energy cost.
We study a system of Brownian quasiparticles moving in a spatially varying temperature landscape containing multiple local minima, see 
Fig.~\ref{fig:1}. 
We show that a swarm of quasiparticles can find the global minimum more efficiently than individual ones. Cooperative clustering emerges in an optimal regime of interaction strength and swarm size that enables reliable identification and dynamic tracking of the global minimum in a time-varying temperature landscape.

 In Sec.~\ref{section:II}, we introduce the physical model of interacting Brownian quasiparticles and the corresponding coarse-grained lattice representation. Our main results are presented in Sec.~\ref{section:III} and in Sec.~\ref{section:IV}. We assess the swarm’s ability to locate the global minimum and identify the regimes of interaction strength and swarm size that yield optimal performance. Furthermore, we investigate the swarm's capability to adapt dynamically to changes in the temperature landscape.
We end our article by discussing the physical interpretation, scalability, and implications for unconventional computing.

\section{System setup and dynamics}
\label{section:II}
In our framework, the swarm agents are realized as interacting Brownian quasiparticles diffusing within a spatially varying temperature field. The interactions between particles and position-dependent thermal noise drive the collective dynamics, enabling the swarm to self-organize and cluster near the global minimum of the temperature profile, as illustrated in Fig.~\ref{fig:1}a). We assume that the time evolution of the quasiparticle position is governed by the overdamped Langevin equations with temperature-dependent multiplicative noise~\cite{Widder1989, Dean1996, Lau2007, Kampen2007, Yang2013}
\begin{equation}
\gamma\partial_t\vect{r}_\alpha(t)= -\sum_{\beta \neq \alpha}^N \vect{\nabla}_{\vect{r}_\alpha} U(\epsilon,\sigma,|\vect{r}_{\alpha} - \vect{r}_\beta
|) + \sqrt{2 \gamma k_B T(\vect{r}_\alpha)} \vect{\eta}_\alpha(t).
\label{eq:langevin}
\end{equation}
Here, $\vect{r}_\alpha(t)$ describes the position of particle $\alpha$ in a square box of length $L$ containing the $N$ quasiparticles.
$\gamma$ represents the damping coefficient, $k_B$ is the Boltzmann constant. We consider a short-range attractive interaction combined with a hard-core repulsion, modeled by the pair potential $U(\epsilon, \sigma, |\vect{r}_{\alpha}-\vect{r}_\beta|)$, where  $\epsilon$ sets the interaction strength and $\sigma \ll L$ denotes the hard-core range. Thermal fluctuations from the environment are simulated as Gaussian white noise $\vect{\eta}_\alpha(t)$ with zero mean and unit variance, scaled by the local temperature $T(\vect{r_\alpha})$. We impose reflective boundary conditions on all sides to prevent particle escape. We monitor the quasiparticles' positions on a uniform grid of $M=\ell^2$ sensors that discretize the two-dimensional square domain of side length $L$, as shown in Fig.~\ref{fig:1}b). Each sensor corresponds to a lattice cell of size $\sigma^2$, with $\sigma=L/\ell$, chosen to match the interaction range. Because of the hard-core repulsion at short distances, each sensor can be occupied by at most one quasiparticle at a time. We denote the resulting discrete configuration of the system by the occupation vector $\vect{\rho} = (\rho_1, \rho_2, \dots, \rho_M)$, where $\rho_s \in \{ 0,1\}$ indicates whether site $s$ is empty or occupied. In this coarse-grained representation, the system state is fully specified by $\vect{\rho}$, and its stochastic dynamics can be described as random transitions between occupation configurations. The exclusion constraint, combined with nearest-neighbor interactions, gives rise to the effective configurational energy
\begin{equation}
\label{eq:config_energy}
    U_\mathrm{eff}(\vect{\rho})=  \epsilon \sum_{\langle s, s^\prime\rangle} \rho_s \rho_{s^\prime},
\end{equation}
where $\langle s, s^\prime\rangle$ represents the nearest-neighbor sum. We assume that the total number of quasiparticles is conserved during the dynamics, i.e. $\sum_{s=1}^{M} \rho_s =N$. When a particle hops from site $s$ to a neighboring site $s^\prime$, the configuration changes from $\vect{\rho}$ to $\vect{\rho}^{s \to s^\prime}$, leading to an energy difference $\Delta U_\mathrm{eff}^{s \to s^\prime}=U_\mathrm{eff}(\vect{\rho}^{s \to s^\prime})-U_\mathrm{eff}(\vect{\rho})$. 
The coarse-grained dynamics of the system is modelled by a continuous-time Markov process where the transition rate between the configurations $\vect{\rho}$ and $\vect{\rho}^{s \to s^\prime}$~\cite{Glauber1963, Seifert2012} is derived from the underlying Langevin equation in Eq.~\eqref{eq:langevin} as
\begin{equation}
\label{eq:transition_rate}
    \Gamma ^{s \to s^\prime}=\frac{k_B\bar{T}_{s,s^\prime}}{\gamma \sigma^2}  \rho_s(1-\rho_{s^\prime}) \sqrt{\frac{T_{s^\prime}}{T_s}} \exp \left[- \frac{\Delta U_\mathrm{eff}^{s \to s^\prime}}{2k_B\bar{T}_{s,s^\prime}}\right].
\end{equation}
Here, $\bar{T}_{s, s^\prime}$ is the harmonic average of the temperature at the two sites $T_s$ and $T_{s'}$ (see App.~\ref{App:rates} for details). Based on Eq.~\eqref{eq:transition_rate}, stochastic trajectories $\vect{\rho}(t)$ are generated using the Gillespie algorithm~\cite{Gillespie1976,Gillespie1977} (see App.~\ref{app:gillespie}). A representative snapshot of $\vect{\rho}(t)$ is shown in Fig.~\ref{fig:1}b), illustrating the instantaneous quasiparticle occupations on the lattice. This corresponds to how occupations would be measured experimentally, where an array of spatially resolved sensors records the system configuration at a given instant.

    To demonstrate the working principle of the proposed optimization scheme, we consider a temperature landscape featuring multiple local minima and a unique global minimum at sensor site $s_0$ (see App.~\ref{app:temperature_profile_parameters} for details of the considered temperature profiles). 
    The central idea is that the quasiparticle swarm collectively identifies the global optimum by clustering around it. We investigate how this behaviour depends on two independent control parameters: the filling fraction $\nu = N/M$ and the dimensionless interaction strength $\epsilon/(k_B T_0)$ between quasiparticles, where $T_0$ denotes the minimum temperature.

In Sec.~\ref{section:III}, we consider a static temperature profile. We initialize the system from a random quasiparticle configuration.
We discard an initial interval of duration $1000\,\tau_0$ to eliminate transient dynamics and redefine the time origin such that $t=0$ corresponds to this point. We then vary the observation time window to determine the minimal duration required to reliably evaluate the performance of the system.
In Sec.~\ref{section:IV}, we instead vary the temperature profile in time. Guided by the results of Sec.~\ref{section:III}, we analyze the time-dependent response of the system by evaluating observables using a sliding time-window averaging scheme.

\begin{figure*}[t]
  \centering
  \includegraphics[width=1\textwidth]{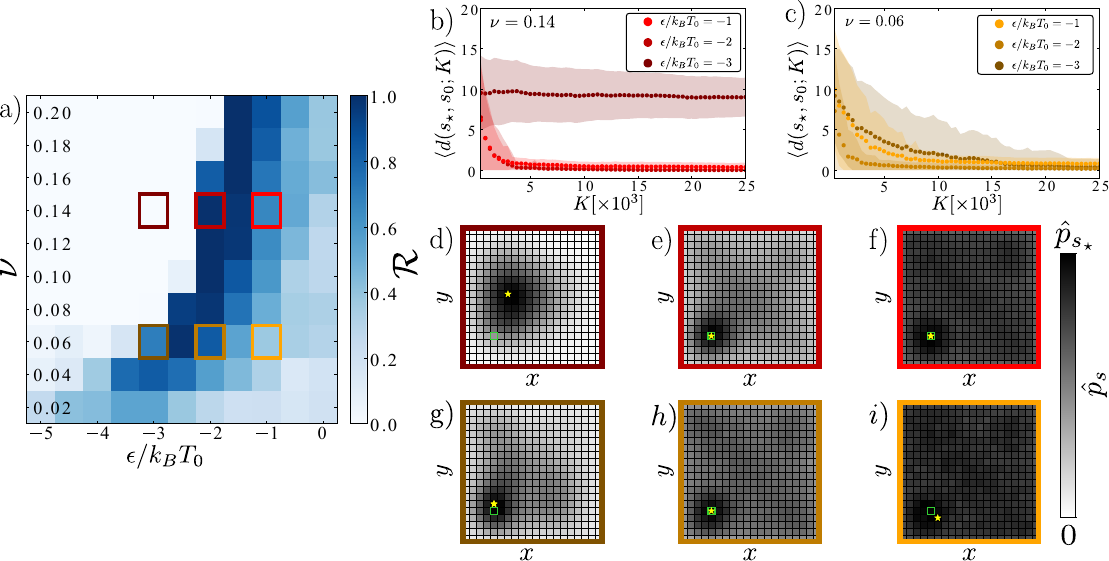}
  \caption{Computational performance of  swarm-based Brownian computing as a function of the dimensionless interaction strength $\epsilon/k_B T_0$ and filling fraction $\nu = N/M$. 
  a) Ensemble-averaged success ratio $\mathcal{R}$. 
  Colored squares mark the parameter combinations illustrated in detail in panels the other panels. 
  (b)--(c) Manhattan distance of the mode position and the position of the global temperature minimum as a function of the time-averaging length characterized by $K$. Dots show the average, and the shaded background color indicates one standard deviation. 
  (d)--(i) Representative steady-state probability distributions  $\hat{\vect{p}}$ for $K=25.000$ and selected parameters $\epsilon/k_B T_0$ and $\nu$ indicated in (a). As in Fig.~\ref{fig:1}, the green box indicates the global temperature minimum and the yellow star the position $s_*$ of the statistical mode of the probability distribution.  
}
  \label{fig:2}
\end{figure*}

\section{Computational Performance}
\label{section:III}

To evaluate the computational performance of the quasiparticle swarm for a static temperature profile, we compare the lattice site with the highest occupation probability $s_*$ to the true temperature minimum $s_0$: 
To this end, we first calculate the single-particle probability distribution $\hat{\vect{p}}$ by time averaging the occupation vector $\vect{\rho}(t)$. 
The average is taken over a time window of duration $K\Delta t$ with $K$ snapshots sampled uniformly in time steps of $\Delta t$. We choose $\Delta t = 4 \tau_0$, i.e.\ a time step longer than the characteristic relaxation time of the quasiparticles $\tau_0=\gamma\sigma^2/(k_B T_0)$ to avoid oversampling. 
Hence, the estimated probability distribution at observation time $J\Delta t$ with $J > K$ is given by
\begin{equation}
    \hat{\vect{p}}(J, K)= \frac{1}{N K}  \sum_{k=0}^{K-1}
    \vect{\rho}\left((J- k) \Delta t\right).
    \label{eq:probability_estimate}
\end{equation}
For an example of the single-particle probability distribution, see Fig.~\ref{fig:1}c).

For sufficiently long sampling times, this temporal average converges to a steady-state probability distribution.
The solution is then identified by the mode of the estimated distribution $\hat{\vect{p}}$, corresponding to the lattice site with the highest occupation probability as $s_\star=\text{argmax}_s \left(\hat{p}_s\right)$, marked by a yellow star in Fig.~\ref{fig:1}c).
A computation is successful when the mode $s_\star$ coincides with the position of the global minimum, i.e.\ when $s_\star~=~s_0$. 

To quantify the reliability of the computation across independent realizations of the simulation, we introduce the success ratio 
\begin{equation}
\label{eq:r}
\mathcal{R}  = \langle \delta_{s_\star, s_0}\rangle,
\end{equation}
as a metric, where $\langle \dots \rangle$ represents averaging over independent simulation realizations. The success ratio $\mathcal{R}$ is the fraction of realizations where the mode of the estimated probability distribution correctly identifies the global minimum.

Fig.~\ref{fig:2} summarizes our analysis across different filling fractions $\nu$ and interaction strengths $\epsilon$.
In the static setting considered here, the probability distribution is evaluated by averaging over the entire observation time, i.e., $J = K$, so that convergence of $\hat{\vect{p}}$ is assessed by increasing $K$.
Fig.~\ref{fig:2}(a) shows the dependence of the success ratio $\mathcal{R}$ on these parameters.
In Figs.~\ref{fig:2}b) and c), we plot the Manhattan distance~\footnote{The Manhattan distance between two lattice points is evaluated as the sum of their horizontal and vertical separation.} $d(s_\star,s_0; K)$ between the mode $s_\star$ and the global minimum $s_0$, as a function of the time-averaging window $K$ for selected representative parameter values. We observe that, for some parameter choices, the Manhattan distance decreases and converges to zero (with varying convergence rates), indicating successful identification of the global minimum. In contrast, for other parameter values (e.g., $\epsilon/k_B T_0=-3$ and $\nu=0.14$), the distance saturates at a finite value.
Figs.~\ref{fig:2}(d–i) show representative steady-state probability distributions, together with their modes (yellow stars) and the site of the global minimum (green boxes).

For the parameters considered in Fig.~\ref{fig:2}(d), the mode remains trapped at a local minimum. By contrast, in Figs.~\ref{fig:2}(e), (f), and (h), the mode correctly converges to the global minimum, while in Figs.~\ref{fig:2}(g) and (i) it converges to a neighboring site. 
Physically, the filling fraction controls how many quasiparticles are available to explore the temperature landscape, whereas the interaction strength determines the strength of their collective behaviour. Increasing the filling fraction enhances spatial coverage, whereas stronger interactions promote clustering and localization. 
At very low filling fractions and weak interactions, the quasiparticles fail to explore the landscape efficiently. Conversely, at high filling fractions and strong interactions, excessive clustering effectively reduces the system’s degrees of freedom, leading to a marked decline in performance. Optimal performance is achieved in an intermediate regime, where interactions assist collective exploration and facilitate reliable identification of the global minimum.

\begin{figure*}[t]
  \centering
  \includegraphics[width=\textwidth]{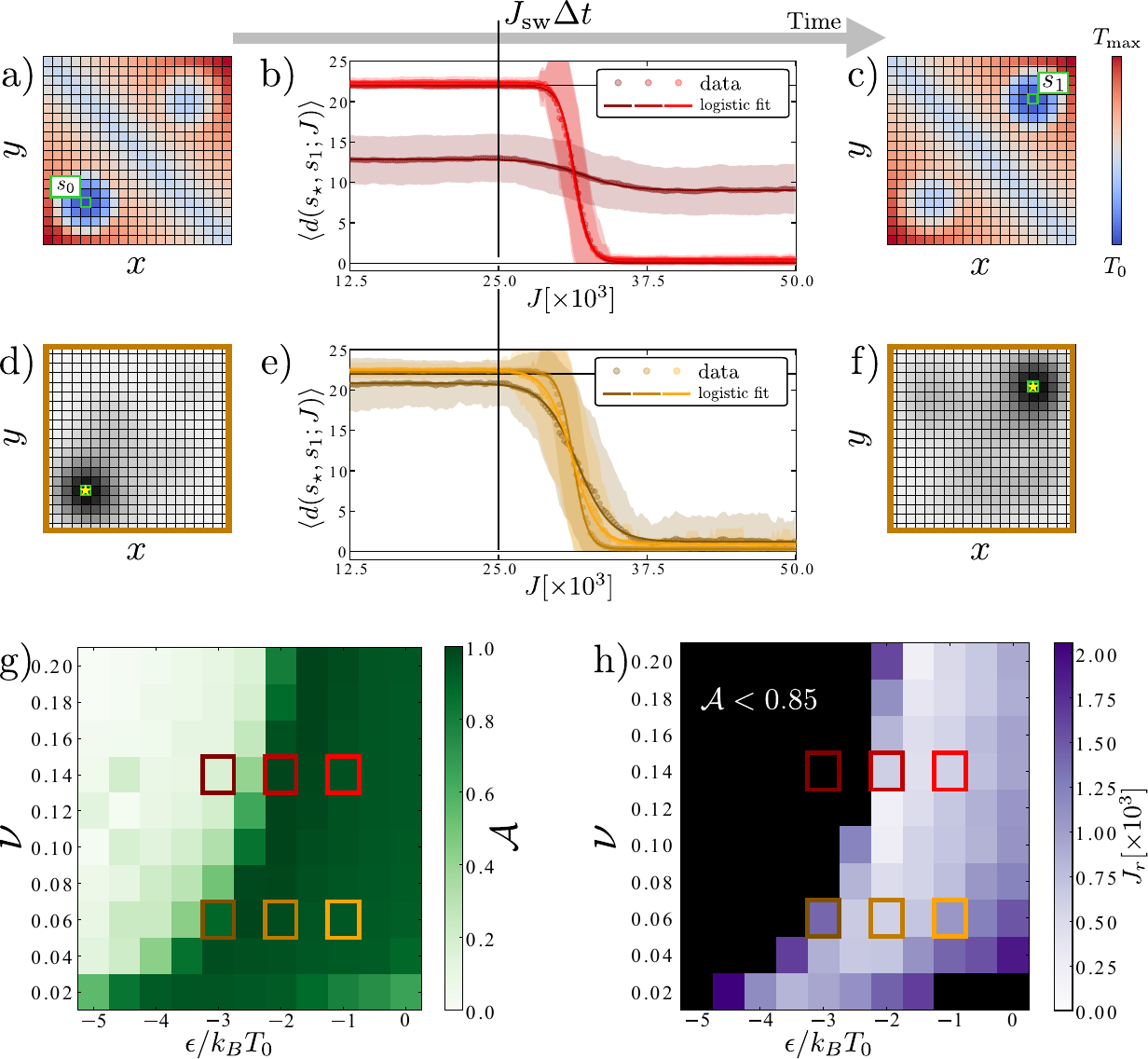}
  \caption{
  Dynamic computational performance of swarm-based Brownian computing as a function of the dimensionless interaction strength $\epsilon/k_B T_0$ and filling fraction $\nu = N/M$.
  (a, c) Temperature landscapes before and after the instantaneous switch of the global minimum from site $s_0$ to $s_1$ at time $J_{\text{sw}}\Delta t$.
  Panels (d) and (f) show representative steady-state probability distributions $\hat{\vect{p}}$ for $\epsilon/k_B T_0=-2$ and $\nu=0.06$, before and after the switch.
  (b, e) Time evolution of the ensemble-averaged Manhattan distance $\langle d(s_\star,s_1;J)\rangle$ between the distribution mode $s_\star$ and the target minimum $s_1$ for different values of $\nu$ and $\epsilon/k_B T_0$ corresponding to the color-coded squares in panels (g) and (h). Shaded regions indicate error bars with one standard deviation.
  Solid lines represent the logistic fit, see Eq.~\eqref{eq:fit}.
 (g) Adaptation accuracy $\mathcal{A}$, defined in Eq.~\eqref{eq:accuracy}, quantifying the success of the transition from the initial to the final global minimum. 
  For the adaptation timescale $J_r$ we only show results for $\mathcal{A}>0.85$ in panel h). 
  }
  \label{fig:switch_mode}
\end{figure*}

\section{Evaluating Dynamic Performance}
\label{section:IV}
We now investigate the ability of the quasiparticle swarm to adapt to nonstationary temperature landscapes. We show that it can track a moving global minimum without requiring a system reset and quantify the associated relaxation dynamics. 
We consider a time-dependent temperature profile in which the location of the global minimum shifts instantaneously from
$s_0$ to $s_1$ at time $t_{\rm sw}= J_{\rm sw}\Delta t$, as illustrated in Fig.~\ref{fig:switch_mode}.
To characterize the system’s dynamical behavior, we employ a sliding time-window average. We fix the averaging window size to $K=25{,}000$ and vary the observation time $J\Delta t$, which allows us to track the temporal evolution of the system.
Representative probability distributions before and after the switch are shown in Figs.~\ref{fig:switch_mode}(d) and (f) for $\epsilon/k_B T_0=-2$ and $\nu=0.06$. For suitable parameter choices, the mode of the distribution successfully relocates to the new global minimum. Figures~\ref{fig:switch_mode}(b) and (e) show the time evolution of the Manhattan distance $d(s_\star,s_1;J)$ between the mode $s_\star$ and the post-switch global minimum $s_1$, for the same combinations of filling fraction and interaction strength considered in the previous section.

In our example, the Manhattan distance $d(s_0,s_1)$ between the initial and final global minimum, $s_0$ and $s_1$, is $d(s_0,s_1)=22$ lattice units, which sets the initial value of $d$ for parameter sets that successfully identify $s_0$. Following the switch, the averaged distance relaxes to zero, indicating that the dominant mode shifts to the new global minimum. This relaxation is not instantaneous: the finite averaging window introduces an intrinsic delay, and the quasiparticle swarm itself requires finite time to adapt to the modified temperature landscape. By contrast, in the high-filling, strong-interaction regime, where performance is already poor in the static case, the system fails to track the new minimum (dark red curve).

To further quantify the ability of the quasiparticle swarm to adapt to the temperature profile change, we introduce an accuracy metric $\mathcal{A}$
measuring the separation between the pre- and post-switch solutions 
and define the adaptation time $J_{\mathrm{ad}}\Delta t$ required to reach the new solution as follows:
We fit the realization-averaged Manhattan distance with a logistic function as,
\begin{equation}
    \langle d(s_\star, s_1; J) \rangle \approx d_i+\frac{d_f-d_i}{1+\exp \left[ \frac{J_m-J}{J_{ad}}\right]}.
    \label{eq:fit}
\end{equation}
Here, the parameters $d_i$ and $d_f$ represent the asymptotic values of the initial and final distance before and after adaptation, respectively, while $J_m$ marks the midpoint of the transition. 
Based on this fit, we extract the parameter $J_{ad}$ which characterizes the timescale associated with convergence to the new solution. 
The accuracy metric $\mathcal{A}$ we define as
\begin{equation}
\label{eq:accuracy}
    \mathcal{A} = 1 - \frac{|d_i-d(s_0,s_1)| + d_f}{d(s_0,s_1)},
\end{equation}
which equals unity in the case of perfect adaptation. 

Figures~\ref{fig:switch_mode}(g) and \ref{fig:switch_mode}(h) show the adaptation accuracy $\mathcal{A}$ and characteristic timescale $J_r$ for different filling fractions and interaction strengths.
Analogous to Fig.~\ref{fig:2}(a), we identify a region in parameter space where convergence is rapid (small $J_r$) and adaptation accuracy is high ($\mathcal{A} \simeq 1$)~\footnote{
The nonidentical parameter dependence of the success ratio $\mathcal{R}$ and the accuracy measure $\mathcal{A}$ reflects the different averaging procedures used to evaluate these quantities.
}. 
These results highlight the existence of an intermediate regime of filling fraction and interaction strength in which the system not only reliably identifies the global minimum in the static case but also efficiently adapts to changes in the temperature landscape.

\section{Discussion}

The computational performance of the interacting quasiparticles emerges from a competition between system exploration and clustering. 
At low filling fractions or weak interactions, quasiparticles, as expected, explore the temperature landscape inefficiently, resembling independent random searchers.
At high filling fractions and strong interactions, excessive clustering reduces the effective degrees of freedom of the quasiparticles, leading to trapping in suboptimal configurations. Optimal performance arises in an intermediate regime where interactions enhance collective exploration without suppressing mobility, enabling both reliable identification of the global minimum and rapid adaptation to changes in the landscape.

The overdamped Langevin dynamics (described by Eq.~\eqref{eq:langevin}) underlying our analysis provides a generic description applicable to a wide range of physical systems obeying  `quasiparticles' in the form of particle-like excitations such as magnetic skyrmions~\cite{Schutte2014, Rozsa2016, Zazvorka2019, Litzius2020, Zhao2020, Weisenhofer2021, Koraltan2026}, superconducting vortices~\cite{Blatter1994, Xu2011}, and other topological solitons but also to real particles like magnetic nanoparticles~\cite{Garcia1998, Xin2025}, colloidal suspensions~\cite{Liu2023, Fortulan2025}, and active biological systems~\cite{Gronbech1994, Julicher1997, Liu2025}. This suggests that the optimization principles demonstrated here are not tied to a specific physical realization.

The collective dynamics studied here fall within the Hohenberg--Halperin Model~B class~\cite{Hohenberg1977}, with the occupation field $\vect{\rho}$ acting as a locally conserved scalar order parameter whose relaxational, diffusive dynamics are driven into a nonequilibrium steady state by the spatially varying temperature field. 
Furthermore, our framework also admits a particle--hole symmetry interpretation, in which empty lattice sites act as effective agents exploring an inverted temperature landscape, thereby enabling the identification of maxima rather than minima.
More generally, while we focus on temperature profiles here, the concept of swarm-based Brownian computing extends to any field to which the quasiparticles are sensitive.

In our model, the effective spatial resolution of the computation is set primarily by the sensor size, rather than by the intrinsic dimensions of the quasiparticles themselves. Depending on the experimental platform, the sensors may, for example, correspond to pixels in optical imaging, magnetoresistive sensor arrays in magnetic systems, or microfluidic compartments in active matter systems~\cite{Crocker1996, Wu2011, Guang2023, Zhao2024}. Sensors much larger than the interaction range $\sigma$ produce an overly coarse description that averages out local temperature gradients and interaction-induced correlations. Conversely, sensors much smaller than the effective particle size lead to redundant discretization, with strongly correlated neighboring sensors providing little additional physical information~\cite{Geyer1992, Liu2016}.

\section{Conclusion and Outlook}
We have shown that interacting Brownian quasiparticles can perform efficient, scalable optimization by exploiting emergent collective behavior driven by short-range interactions and thermal fluctuations. Using a spatially varying temperature landscape as a benchmark optimization task, we demonstrated that interacting swarms reliably identify global optima and outperform non-interacting searchers within a well-defined regime of interaction strength and filling fraction. 
Beyond steady-state optimization, we showed that the swarm can dynamically adapt to nonstationary landscapes by tracking a shifting global minimum without requiring a system reset, consistent with the same underlying trade-off between exploration and clustering.

While our analysis is phrased in terms of quasiparticles intrinsic to the material—where increasing swarm size incurs only minimal energetic cost—the same principles apply to swarms of real particles, albeit with higher energetic overheads typical of engineered multi-agent systems.
A key advantage of our approach is that the swarm agents are quasiparticles intrinsic to the material, so that increasing the swarm size incurs only minimal energetic cost. 
As a result, the approach remains energy-efficient with increasing system size and task complexity within the optimal parameter regime. 
While the same optimization principles may also apply to swarms of real particles, such implementations would typically incur substantially higher overheads due to additional hardware demands, increased engineering complexity, and the greater control energy required.
This intrinsic, material-level scalability, together with the demonstrated robustness to time-dependent inputs, positions interacting Brownian quasiparticles as a promising route toward energy-efficient, physics-based unconventional computing.
Future work may extend this framework to continuous landscapes, incorporate long-range or anisotropic interactions, and explore concrete experimental realizations in materials.

\section{Acknowledgements}
We thank Alfred Hucht and Philipp Gessler for helpful discussions. We also thank Björn Dörschel for his assistance during the initial phase of the project.
We acknowledge funding
from the German Research Foundation (DFG) Project-ID 405553726 (CRC/TRR 270, project B12) and 403233384 (SPP Skyrmionics).

\appendix

\section{From the continuous overdamped Langevin equation to lattice dynamics}
\label{App:rates}
In this Appendix, we motivate Eq.~\eqref{eq:transition_rate}, 
translating the continuous description of particle motion described by the overdamped Langevin equation into a lattice-based, discrete formulation based on transition rates.
In the discretized description, we consider the hopping of a quasiparticle between the midpoints of lattice sites $s$ and $s'$. The ratio of the forward and backward transition rates between these two sites takes the form~\cite{vanKampen1988, Kampen2007, Lau2007, Seifert2012}
\begin{equation}
\label{eq:ratio_rates}
\frac{\Gamma^{s\to s^\prime}}{\Gamma^{s^{\prime}\to s}} = 
\exp \left[\,\int_{\vect{r}_s}^{\vect{r}_{s^\prime}} d\vect{r} \;\frac{-\vect{\nabla}U(\vect{r})+k_B\vect{\nabla}T(\vect{r})}{T(\vect{r})}\right],
\end{equation}
where $\vect{r}_s$ and $\vect{r}_{s^\prime}$ are the midpoints of the two lattice sites, respectively. In our model, we restrict dynamics to nearest-neighbor hops. 
The corresponding line integral is therefore taken over the common boundary of two lattice sites, where in the integral over the first term we approximate the temperature $T(\vect{r})$ by the harmonic average $\bar{T}_{s,s'}$ of $T(\vect r_s)$ and $T(\vect r_{s'})$ in addition to the same approximation for the interaction potential as in the main text. The second integral is calculated exactly. This yields the transition rate ratio 
\begin{align}
\label{eq:appendix_transition_rate_final}
\frac{\Gamma^{s\to s^\prime}}{\Gamma^{s^\prime\to s}} & \simeq \frac{T_s}{T_{s^\prime}} \exp\left[-\frac{\Delta U_{\textrm{eff}}^{s\to s^\prime}}{\bar{T}_{s,s'}} \right],
\end{align}
which is consistent with Eq.~\eqref{eq:transition_rate}. 
Furthermore, the prefactor $k_B\bar{T}_{s,s'}/(\gamma \sigma^2)$ in Eq.~\eqref{eq:transition_rate}  ensures consistent diffusion with the Langevin description, and the factor $\rho_s(1-\rho_{s'})$ ensures hopping from filled to empty spaces only.
The factor $\sqrt{T_s/T_{s'}}$ originates from the symmetric decomposition of the rate ratio $\Gamma^{s\to s'}/\Gamma^{s'\to s} \propto T_s/T_{s'}$, which ensures that forward and backward transitions are treated equivalently.

\section{Generation of stochastic trajectories by means of the Gillespie algorithm}
\label{app:gillespie}
To simulate the stochastic dynamics of the interacting Brownian quasiparticles, we employ a rejection-free Gillespie algorithm~\cite{Gillespie1976, Gillespie1977}. This method generates statistically exact trajectories of a continuous-time Markov process governed by the transition rates defined in Eq.~\eqref{eq:transition_rate}. Starting from a given configuration $\vect{\rho}(t)$, the Gillespie algorithm proceeds as follows:

\begin{enumerate}
    \item{\textit{Enumeration of the possible transitions}:} All possible nearest-neighbor hops $s \to s'$ are identified, that satisfy the exclusion constraint $\rho_s=1, \rho_{s'}=0$. The allowed transitions are then indexed by $n$ such that the corresponding rates $\{\Gamma_n\}$ are ordered in increasing magnitude. The total rate at which the system leaves the current configuration is given by $\Gamma_{\mathrm{tot}} = \sum_{n} \Gamma_n$.    
    \item{\textit{Sampling of the transition time}:} The stochastic waiting time until the next transition is sampled as 
    \begin{equation}
        \delta t = -\frac{1}{\Gamma_{\mathrm{tot}}}\ln(p_1),
    \end{equation}    
    following the exponential distribution of waiting times associated with Poissonian escape from a Markov state. Here \(p_1 \in (0,1]\) is a uniform random number and the average waiting $\langle \delta t \rangle$ is proportional to the system's time scale $\tau_0$.
    \item{\textit{Selecting the transition:}} A second uniform random number \(p_2 \in (0,1]\) is generated to determine which of the possible transitions occurs. The selected transition $\tilde{n}$ corresponds to the smallest transition index satisfying \begin{equation}
        p_2\,\Gamma_{\mathrm{tot}} \leq \sum_{n=1}^{\tilde{n}} \Gamma_n.
    \end{equation} This ensures that each transition is chosen with probability $\Gamma_n/\Gamma_{\mathrm{tot}}$. 
    \item {\textit{Updating the system's state:}} The selected transition $\tilde{n}$ is executed, and the configuration is evaluated at time $t+\delta t$.
\end{enumerate}

Note that we chose the observational sensor resolution $\Delta t$ to be larger than the natural time scale $\tau_0$, i.e.\  $\Delta t = 4 \tau_0 \sim  \langle \delta t \rangle$ to balance sampling efficiency with the system's dynamics.

\section{Temperature profile}\label{app:temperature_profile_parameters}
For the simulations, we discretize the temperature field $T(\vect{r})$ by assigning to each of the $20 \times 20$ cells the value of T evaluated at its center. 
As a temperature field, we used
\begin{equation}
T(\vect{r})
 = T_0\,
 \frac{\Theta (\vect r) }
      {\min
      \Theta (\vect r) },
\label{eq:app:temperature_profile}
\end{equation}
where $T_0$ is the minimal temperature, $\vect{r} =(r_1,r_2)$, and 
the profile function $\Theta (\vect r)$ 
comprises a linear function with slope $g$ and two parabolic well minima
\begin{equation}
\Theta (\vect r)=    g\,|r_1+r_2| + w_{d_1,a_1}(\vect{r}-\vect{r}_{w_1}) + w_{d_2,a_2}(\vect{r}-\vect{r}_{w_2}) 
\end{equation}
that are centered at $\vect{r}_{w_1}$ and $\vect{r}_{w_2}$. 
The wells are parametrized by their depth $d$ and characteristic width $a$ as
\begin{equation}
w_{d,a}(\vect{r}) = d\,\max \Bigl(0, 1-\frac{|\vect{r}|^2}{a^2}\Bigr).
\end{equation}
The denominator in Eq.~\eqref{eq:app:temperature_profile}
normalizes the temperature field such that $\min_{\vect{r}} T(\vect{r}) = T_0$.

For the profile shown in Fig.~\ref{fig:1} and the results discussed in Sec.~\ref{section:III} and Sec.~\ref{section:IV} 
we chose $T_0=1$, $g=0.15$, $\vect{r}_{w_1} = -\vect{r}_{w_2}= (-0.3,-0.3)$ with $a_1=a_2=3.0$.
Furthermore, we chose $d_1= 0.2$ and $d_2=0.1$, making the well at $\vect{r}_1$ the global minimum.
In Sec.~\ref{section:IV}, we interchange the values of $d_1$ and $d_2$ at the switching time $J_{\mathrm{sw}}\Delta t$, thereby shifting the position of the global minimum.
The parameters of the temperature field used in the simulations are summarized in Tab.~\ref{tab:temperature_values}.

\begin{table}[h!]
\centering
\label{tab:temperature_values}
\begin{tabular}{|l|l|l|}
\hline
\textbf{Parameter} & \textbf{Symbol}  
& \textbf{Value} \\
\hline
Min. Temperature & $T_0$ 
& 1.0 \\
Slope & $g$ 
& 0.15 \\
Well centers & $\vect{r}_{w_1}=-\vect{r}_{w_2}$ 
& $(-0.3,\,-0.3)$ \\
Well widths & $a_1 = a_2$ 
& 0.2 \\
Initial well depths & $d_1=d_2/2$ 
& 0.2 \\
\hline
\end{tabular}
\caption{Parameters used to model the temperature field $T(\vect{r})$ in Eq.~\eqref{eq:app:temperature_profile} for the results presented in the main text.}
\end{table}

\bibliographystyle{apsrev4-2}
\bibliography{bibliography}

\end{document}